\documentclass[preprintnumbers]{revtex4}

\usepackage[dvipdfmx]{graphicx,color}
\usepackage{epsf}
\usepackage{amsmath,amssymb}
\usepackage{bm}
\usepackage{textcomp}
\newcommand{\bvec}{\boldsymbol}
\begin{document}
\preprint{KUNS-2541}
\title{Entanglement entropy and Schmidt number as measures of 
delocalization of $\alpha$ clusters in one-dimensional
nuclear systems}
\author{Yoshiko Kanada-En'yo}
\affiliation{Department of Physics, Kyoto University, Kyoto 606-8502, Japan}

\begin{abstract}
We calculated the von Neumann entanglement entropy and the Schmidt number of 
one dimentional (1D) cluster states and showed that these are useful measures to estimate 
entanglement caused by delocalization of clusters. 
We analyze system size dependence of these entanglement measures in the linear-chain 
$n\alpha$ states given by Tohsaki-Horiuchi-Schuck-R\"opke wave functions for 
1D cluster gas states. We show that
the Schmidt number is an almost equivalent measures to the von Neumann entanglement entropy 
when the delocalization of clusters occurs in the entire system but it shows 
different behaviors in a partially delocalized state containing localized clusters and delocalized ones. 
It means that the R\'enyi-2 entanglement entropy, which relates to the Schmidt number, is found to be almost equivalent 
to the von Neumann entanglement entropy for the full delocalized cluster system 
but it is less sensitive to the partially delocalized cluster system than 
the von Neumann entanglement entropy.
We also propose a new entanglement measure which has a generalized form of 
the Schmidt number. Sensitivity of these measures of entanglement to 
the delocalization of clusters in low-density regions was discussed.
\end{abstract}

\maketitle

\section{Introduction}
Nuclear many-body systems are self-bound systems of four species of 
Fermions, spin up and down protons and neutrons.
There, a variety of phenomena arise originating in many-body correlations. 
One of the remarkable spatial correlations is "cluster" which is a subunit composed of 
strongly correlating nucleons. A typical cluster in nuclear systems is 
an $\alpha$ cluster, which is a composite particle consisting of four nucleons.
If there is no correlation between nucleons in a nucleus, 
all nucleons behave as independent particles and the nucleus 
is an uncorrelated state with a clear Fermi surface written by a single Slater determinant
wave function. However, in reality, correlations between nucleons are rather strong and
$\alpha$ clusters are often formed at the surface in particular in light nuclei.
Once $\alpha$ clusters are formed, 
delocalization of clusters occurs to gain the kinetic energy of 
center of mass motion (c.m.m.) of clusters in some situations. 
The delocalization of clusters involves many-body correlations beyond a Slater determinant.

In realistic nuclear systems, a degree of the (de)localization of clusters depends on 
the competition (balance) of the kinetic energy and potential energy of clusters 
and is strongly affected also by the Pauli blocking between nucleons in
clusters and a core nucleus.  
The delocalization limit of $\alpha$ clusters is an $\alpha$ cluster gas state, 
where all $\alpha$ clusters move almost freely like a gas. 
Such a cluster gas has been predicted to appear in the second $0^+$ state of 
$^{12}$C \cite{uegaki1,Tohsaki:2001an}. 
To describe cluster states of delocalized $\alpha$ clusters,   
a new type of cluster wave function, the
so-called ``Tohsaki-Horiuchi-Schuck-R\"opke'' (THSR) wave function, has been introduced
by Tohsaki {\it et al.} \cite{Tohsaki:2001an}. 
The THSR wave function is essentially based on $\alpha$ clusters in a common Gaussian orbit 
having a range of the system size, and suitable to describe general cluster gas states 
of $n$ $\alpha$ clusters. Indeed, it has been shown that $^8$Be($0^+_1$) and $^{12}$C($0^+_2$) 
can be described well by the THSR wave functions of $2\alpha$ and $3\alpha$, 
respectively.  Since the optimized THSR wave functions for these states have 
much larger ranges of the Gaussian orbit of clusters than the cluster size,
 $^8$Be($0^+_1$) and $^{12}$C($0^+_2$)  are interpreted as 
gas-like cluster states of $2\alpha$ and $3\alpha$ \cite{Tohsaki:2001an,Funaki:2002fn,Funaki:2003af,Funaki:2009zz}.

The THSR wave function has been extended to apply to $^{20}$Ne, 
and found to be able to describe also 
$^{16}$O+$\alpha$ states in $^{20}$Ne \cite{Zhou:2012zz,Zhou:2013ala}.
Recently, Suhara {\it et al.} have proposed that 
this concept of the $\alpha$-cluster gas is applicable also to one dimension (1D) cluster motion
in  linear-chain $n\alpha$ structures \cite{Suhara:2013csa}. 
They have proposed the 1D-THSR wave functions and shown that 
the 1D-THSR wave functions with the optimized Gaussian ranges 
can describe well the exact solutions of the linear-chain 
$3\alpha$ and $4\alpha$ states. Their result somewhat contradicts to the 
the conventional picture of the linear-chain $n\alpha$ structures that  
spatially localized $\alpha$ clusters are arranged in 1D with certain 
intervals \cite{Morinaga:1956zza}. 
Even though their model restricted in 1D is not enough to settle
the problem of stability of the linear-chain states in realistic nuclear systems in 3D,
their work provides a new picture of 1D cluster states and may lead to 
a  understanding of cluster phenomena 
in nuclear many-body systems. Moreover, the 1D cluster state is 
an academically interesting problem of quantum many-Fermion systems.

To distinguish between localization and delocalization of composite particles (clusters)
in microscopic wave functions of Fermion (nucleon) systems,
one should carefully consider the antisymmetrization effect of nucleons between clusters. 
When clusters largely overlap with each other, 
motion of clusters is strongly affected by the Pauli blocking between nucleons in other clusters. 
As a result of Pauli blocking effect, when the system size is as small as or smaller than the cluster size, 
clusters can not move freely and the system becomes equivalent to a localized cluster state. 
In the case of a large system size, where the overlap between clusters is small,  clusters can 
move rather freely. It means that the delocalization occurs not in high-density regions but in
low-density regions. Indeed, we have investigated the cluster motion of an $\alpha$ cluster around the 
$^{16}$O and shown that the delocalization of the cluster occurs in a long tail part 
far from the core \cite{Kanada-En'yo:2014vva}. 
Also in the 1D $n\alpha$ states, Suhara {\it et al.} have shown 
peak structures in the density distributions of the linear-chan $3\alpha$ and $4\alpha$ states
suggesting partial localization at high density inner regions even in the 1D gases of $3\alpha$ and $4\alpha$.

To estimate many-body correlations caused by the delocalization of clusters
we need a measure of correlations that is free from the antisymmetrization effect.
There are several entanglement measures that are essentially based on density matrix
and can be used to estimate many-body correlations in quantum many-body systems. 
The Schmidt number \cite{grobe94} is one of the measures, and the von Neumann entanglement entropy \cite{bennett96} is another measure.  
The former has been proposed by Grobe {\it et al.} in 1994 \cite{grobe94}, 
and applied to measure entanglement, i.e., many-body 
correlations in such systems as atomic and nuclear systems 
(see, for example, Refs.~\cite{law05,Sandulescu:2008qv,tichy11} and references therein).  
The latter has been introduced by Bennett {\it et al.} in 1996 \cite{bennett96}, and widely
used in various fields such as condensed matter and 
quantum field theory (see 
Refs.~\cite{Calabrese:2004eu,Plenio07,amico08,Horodecki09,Nishioka:2009un} and references therein). 
In the previous paper, we have calculated the entanglement entropy for the one-body 
density matrix in 1D cluster states 
to measure the entanglement (correlations) caused by the delocalization of clusters, 
and found that the delocalization occurs in low-density regions but is relatively 
suppressed in high density regions \cite{enyo-ee}.

Let us consider uncorrelated and correlated states of Fermion systems.
An uncorrelated state is given by a Slater determinant wave function, for which 
the one-body density matrix $\hat\rho$ is a projector $\hat\rho^2=\hat\rho$ 
in the single-particle Hilbert space. It means that the 
eigen values $\rho_l$ of one-body density matrix satisfies $\rho_l^2=\rho_l$, 
i.e., $\rho_l=1$ or 0.
$\rho_l$ is the occupation probability of the single-particle 
basis that diagonalizes the one-body density matrix and it is nothing but the 
Schmidt coefficients in the Schmidt decomposition for the one-body density matrix.
Both of the Schmidt number and entanglement entropy are entanglement measures which 
can estimate how the Schmidt coefficients deviates from the condition $\rho_l^2=\rho_l$
for uncorrelated states. However, these measures have different dependences on 
the one-body density matrix, i.e., the eigen values $\rho_l$. 

In this paper, we calculate the entanglement entropy and the Schmidt number of 
1D cluster states and make it clear whether these entanglement measures  are 
useful to investigate the delocalization of clusters. 
We analyze the system size dependence of these measures in 1D $\alpha$-cluster states 
given by the 1D-THSR wave functions. We show that
the entanglement entropy and the Schmidt number are almost equivalent measures
when the delocalization of clusters occurs in the entire system but they show
different behaviors in partially delocalized cluster states composed of 
localized clusters and delocalized ones. 
We also propose a new entanglement measure which has a generalized form of 
the Schmidt number. 

This paper is organized as follows. 
We describe the entanglement  measures defined by the one-body density matrix
in Section \ref{sec:formulation}. 
Section \ref{sec:toy} discusses properties of these measures 
of ideal states in a toy model.
In section \ref{sec:1D}, we calculate these measures in 1D nuclear systems of $\alpha$ clusters 
and discuss sensitivity of these measures to the delocalization of clusters. 
The paper concludes with a summary in section \ref{sec:summary}.

\section{Measures of entanglement}\label{sec:formulation}
The entanglement entropy and Schmidt number are the measures 
of entanglement in quantum many-body systems, which are defined by the density matrix. 
In this section, we describe these measures and also propose 
a new measure by extending the Schmidt number. We also define
spatial distributions of these  entanglement measures.

\subsection{One-body density matrix}
In the present paper, we use only the one-body density matrix, 
which has been often discussed in nuclear systems \cite{ring-schuck},
though more general density matrices are used to define measures of 
entanglement. 
For a wave function $|\Psi^{(A)}\rangle$ of an $A$-particle state, 
the matrix element of the one-body density in an arbitrary  
basis 
is given as
\begin{equation}
\rho_{pq}=\langle\Psi^{(A)}|c^\dagger_q
c_p|\Psi^{(A)} \rangle, 
\end{equation} 
where $c^\dagger_p$ and $c_p$ are the creation and annihilation operators, respectively. 
The one-body density matrix is regarded as the matrix element of the one-body density 
operator $\hat \rho_{\Psi}$ for the wave function $\Psi^{(A)}$, 
\begin{equation}
\hat \rho_{\Psi}|=\sum_{pq} |p\rangle \rho_{pq} \langle q|.
\end{equation} 
The one-body density matrix can be diagonalized by a unitary transformation of 
single-particle basis
\begin{eqnarray}
(D^\dagger \rho D)_{ll'}&=&\rho_l \delta_{ll'},\\
a^\dagger_l&=&\sum_{l'} D_{l'l} c^\dagger_{l'},
\end{eqnarray}
where
\begin{eqnarray}
\rho_l&=&\langle\Psi^{(A)}|a^\dagger_l a_l  |\Psi^{(A)}\rangle,\\
0&\leq& \rho_l \leq 1
\end{eqnarray}
is the eigen value of the one-body density matrix and means the occupation probability of 
the single-particle state $l$ in the wave function $\Psi^{(A)}$.  $\rho_l$ corresponds to
so-called Schmidt coefficients in the Schmidt decomposition for the one-body density matrix.
The trace of the one-body density matrix $\rho$ equals to the particle number $A$ as
\begin{eqnarray}
A&=&{\rm Tr} \rho=\sum_l\rho_l. 
\end{eqnarray}
Note that the normalization of the one-body density matrix is not a unit but 
is the particle number $A$. In the present paper, the entanglement entropy and the Schmidt number
are defined by thus defined one-body density matrix normalized as ${\rm Tr} \rho=A$.

If  a wave function $|\Psi^{(A)}\rangle$ is a Slater determinant, 
$\rho_l=1$ for occupied single-particle states and $\rho_l=0$ for unoccupied single-particle states.
It means that the one-body density operator $\hat \rho_{\Psi}$
satisfies ${\hat{\rho}_{\Psi}}^2=\hat \rho_{\Psi}$ and is a projector 
in the single-particle Hilbert space if $\Psi^{(A)}$ is a non-entangled state 
given by a Slater determinant wave function.

\subsection{Measures of entanglement}
\subsubsection{Entanglement entropy}
The von Neumann entanglement entropy has been introduced by Bennett {\it et al.} and 
proved to be an entanglement measure in quantum many-body systems (see, for instance, Refs.~\cite{tichy11,amico08} and references therein).
The entanglement entropy is defined by the von Neumann entropy of one of the reduced density matrices and called ``von Neumann entanglement entropy'', 
which we call the ``entanglement entropy'' unless otherwise noted. 
In the present paper, we consider the entanglement entropy only for the one-body density matrix
of Fermion systems.   
The entanglement entropy that is defined by the one-body density matrix is given as,
\begin{equation}
S=-{\rm Tr}\rho\log \rho=- \sum_l \rho_l \log \rho_l.
\end{equation}
The entanglement entropy is zero if 
a wave function $|\Psi^{(A)}\rangle$ is a Slater determinant, because 
$\rho_l=1$ for occupied single-particle states and $\rho_l=0$ for unoccupied single-particle states.
It means that a system has non-zero positive value of the 
entanglement entropy only if  the system contains 
many-body correlations beyond a Slater determinant, i.e., if the system is entangled. 

\subsubsection{Schmidt number}
Another entanglement measure is the so-called Schmidt number, which has been introduced 
by Grobe {\it et al.} and used 
to measure many-body correlations in atomic and nuclear physics \cite{law05,Sandulescu:2008qv,tichy11}.
The Schmidt number $K$ is defined as, 
\begin{equation}
K=\frac{A}{\sum_l \rho^2_l}=\frac{A}{\textrm{Tr}\rho^2},
\end{equation}
which estimates the number of states involved in the Schmidt decomposition.
$K$ equals one if a wave function $|\Psi^{(A)}\rangle$ is a Slater determinant, and $K$ is greater than one for entangled states. In analogy to the entanglement entropy, which is generated by the 
entanglement, it is useful to consider  the quantity $K-1$, 
\begin{equation}\label{eq:K-1}
K-1=\frac{A}{\sum_l \rho^2_1}-1= 
\frac{\sum_l (\rho_l-\rho^2_l)}{\sum_l \rho_l^2}.
\end{equation}
It is clear that $K-1=0$ only for non-entangled states 
because $\rho^2_l=\rho_l$, i.e.,  ${\hat \rho_\Psi}^2={\hat \rho_\Psi}$
is satisfied only if a state is given by a Slater determinant. It means that 
a non-zero positive value of $K-1$ is generated by entanglement, that is, many-body correlations beyond a Slater determinant.
One of the merits of the $K$ number is that it is given by 
${\textrm{Tr}\rho^2}$ which can be calculated by the matrix element
$\rho_{pq}$ in an arbitrary basis without the diagonalization of the one-body density matrix.

We should comment that the logarithm of the Schmidt number is 
nothing but the R\'enyi entanglement entropy 
of order 2 (R\'enyi-2 entanglement entropy) for 
the one-body density matrix,
\begin{eqnarray}
\log{K}&=&S^{\rm Renyi}_{2},\\
S^{\rm Renyi}_{\xi}&=&\frac{1}{1-\xi}\log \{\frac{\textrm{Tr}\rho^\xi}{A}\}.
\end{eqnarray}
It is known that, in the $\xi\to 1$ limit, the R\'enyi entanglement entropy 
becomes equal to the von Neumann entanglement entropy. 
In this paper, we discuss the Schmidt number instead of R\'enyi-2 entropy
though they are equivalent entanglement measures.  

\subsubsection{Extension of Schmidt number}
\begin{figure}[htb]
\begin{center}
	\includegraphics[width=5cm]{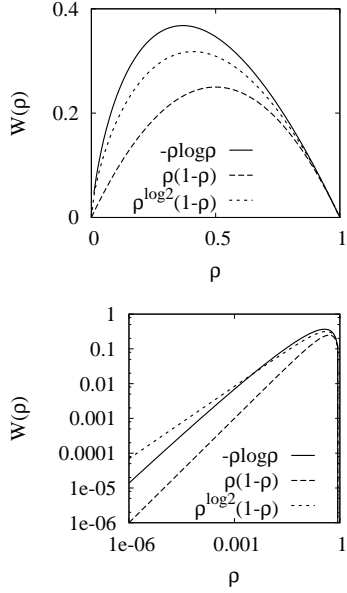} 	
\end{center}
  \caption{Weight functions $W(\rho)=-\rho\log\rho$, $\rho(1-\rho)$, 
and $\rho^{\log 2}(1-\rho)$ for 
$S$, $K$, and $K_{\log 2}$, respectively. 
\label{fig:dens}}
\end{figure}

In the present paper, we propose an entanglement measure which is regarded as a generalized 
version of the Schmidt number. 
As shown in Eq.~\eqref{eq:K-1},  the origin of non-zero contributions in $K-1$ is 
partially occupied single-particle state with $0< \rho_l <1$. 
Ignoring the total scaling factor $1/{\sum_l \rho_l^2}$, 
the contribution $W(\rho_l)$ (the weight function) of a single-particle state 
with the occupation probability $0< \rho<1$ in $K-1$ has the $\rho_l$ dependence,
$W(\rho_l)=\rho_l-\rho^2_l$, which is different from the weight function
$W(\rho_l)=-\rho_l\log \rho_l$ in the entanglement entropy $S$. 
The former has relatively small weights for single-particle states with small occupation probability $\rho_l < 1/2$
compared with the latter although 
the occupied $\rho_l=1$ and unoccupied $\rho_l=0$ single-particle states have no weight
in both cases (see the $\rho$ dependence of the weight functions in Fig.~\ref{fig:dens}).  
We here introduce an alternative weight function, 
$W(\rho_l)=\rho_l^\gamma(1-\rho_l)$, and define an entanglement measure $K_\gamma$
by extending the entanglement measure $K$ as follows, 
\begin{eqnarray}
K_\gamma&=&\frac{\sum_l \rho^\gamma_l}{\sum_l \rho_l^{(1+\gamma)}}
=\frac{\textrm{Tr}\rho^\gamma}{\textrm{Tr}\rho^{(1+\gamma)}},\\
K_\gamma-1&=&\frac{\sum_l \rho^\gamma_l(1-\rho_l)}{\sum_l \rho_l^{(1+\gamma)}},
\end{eqnarray}
where the parameter $\gamma$ is a positive constant.
In the case of $\gamma=1$, the $K_\gamma$ number becomes consistent with the Schmidt number $K$. Similarly to the known entanglement measures ($S$ and $K-1$), 
$K_\gamma-1$ equals to zero only for a Slater determinant, whereas
a non-zero positive value of $K_\gamma-1$ is generated by entanglement, that is, 
many-body correlations beyond a Slater determinant.
Note that the diagonalization of the density matrix is required 
to obtain the $K_\gamma$ number for a non-integer $\gamma$ differently from 
the $K$ number. 
In this paper, we choose $\gamma=\log 2$ which gives the weight function 
$W(\rho_l)=\rho_l^{\log 2}(1-\rho_l)$ having $\rho_l$ dependence similar 
to $W(\rho_l)=-\rho_l\log \rho_l$ for the entanglement entropy in $0.001\lesssim\rho\le1$
(see Fig.~\ref{fig:dens}). 

It should be commented that $\log K_\gamma$ is given by the R\'enyi entanglement entropy 
of order $\gamma$ and $1+\gamma$ as,
\begin{eqnarray}
\log{K_\gamma}&=&(1-\gamma)S^{\rm Renyi}_\gamma+\gamma S^{\rm Renyi}_{1+\gamma}.
\end{eqnarray}

\subsection{Spatial distributions of entanglement measures}

To investigate the spatial distribution of the "important single-particle states" 
that contribute to 
the non-zero entanglement entropy,  we have defined
the spatial distribution $s(\bvec{r})$ of the entanglement entropy, which have been introduced 
in the previous paper \cite{enyo-ee} as, 
\begin{eqnarray}
S&=& \sum_l (-\rho_l\log \rho_l) = \int   s(\bvec{r}) d\bvec{r},\\
s(\bvec{r})&=&\sum_l  \left(-\rho_l\log\rho_l\right)
 \phi^*_l(\bvec{r})\phi_l(\bvec{r}).
\end{eqnarray}
Here the factor $-\rho_l\log\rho_l$ is contribution 
of the single-particle state $|l\rangle$ in $S$, and $\phi^*_l(\bvec{r})\phi_l(\bvec{r})$ means the 
density distribution in $|l\rangle$ and it is normalized as 
$\int \phi^*_l(\bvec{r})\phi_l(\bvec{r})d\bvec{r}=1$.
Therefore, the distribution $s(\bvec{r})$ reflects 
spatial distributions of the important single-particle states $|l\rangle$ that contribute to 
the total entanglement entropy, whereas it is hardly affected by almost occupied single-particle states 
having $\rho_l\approx 1$.  
The expression of $S$ with $s(\bvec{r})$ is analogous to that of the particle number with the
density distribution, 
\begin{eqnarray}
A&=&\sum_l\rho_l=\int\rho(\bvec{r}) d \bvec{r},\\
\rho(\bvec{r})&=&\sum_l \rho_l\phi^*_l(\bvec{r})\phi_l(\bvec{r}).
\end{eqnarray}
Note that the distribution $s(\bvec{r})$ is not quantity 
determined only by local information at the position $\bvec{r}$. It is different from the 
density distribution $\rho(\bvec{r})$ which is determined only by the local information. 

We also define the distributions
$\kappa(\bvec{r})$ and $\kappa_\gamma(\bvec{r})$ 
for $K-1$ and $K_\gamma-1$, respectively, 
to see spatial distributions of the "important single-particle states" in total amount of the measures, 
\begin{eqnarray}
K-1 &=&\frac{\sum_l (\rho_l-\rho^2_l)}{\sum_l \rho_l^2}\nonumber\\
&=& \int   \kappa(\bvec{r}) d\bvec{r},\\
\kappa(\bvec{r})&=&\sum_l    \frac{1}{\sum_l' \rho_l'^2}
(\rho_l-\rho^2_l)
\phi^*_l(\bvec{r})\phi_l(\bvec{r})\nonumber\\
&=&\sum_l    \frac{K}{A}(\rho_l-\rho^2_l)
\phi^*_l(\bvec{r})\phi_l(\bvec{r}),
\end{eqnarray}
and 
\begin{eqnarray}
K_\gamma-1 &=&\frac{\sum_l \rho^\gamma_l(1-\rho_l)}{\sum_l \rho_l^{1+\gamma}}\nonumber\\
&=&\int   \kappa_\gamma(\bvec{r}) d\bvec{r},\\
\kappa_\gamma(\bvec{r})&=&\sum_l  \frac{1}{\sum_{l'} \rho_{l'}^{1+\gamma}} \rho^\gamma_l(1-\rho_l)
\phi^*_l(\bvec{r})\phi_l(\bvec{r}).
\end{eqnarray}
Similarly to $s(\bvec{r})$, these distributions, $\kappa(\bvec{r})$ and 
$\kappa_\gamma(\bvec{r})$, show
spatial distributions of the important single-particle states $|l\rangle$ that contribute to 
the total amount of $K-1$ and $K_\gamma-1$, respectively. 
Note that these distributions are not local quantities. 

\section{$S$, $K$, and $K_\gamma$ of ideal states in a toy model}\label{sec:toy}
In this section, we discuss behaviors of $S$, $K$, and $K_\gamma$ 
for correlated (entangled) states composed of delocalized
clusters in a toy model.
Here, $\gamma$ in the $K_\gamma$ number is assumed to be $0<\gamma<1$.

Let us consider $n_f$ species of particles. For instance, $n_f=2$ for 
spin up and down Fermions, and $n_f=4$ for spin up and down 
protons and neutrons. We use the label $\sigma$ for species of Fermions
such as $\sigma=p\uparrow$, $p\downarrow$, $n\uparrow$, $n\downarrow$
for nuclear systems. 
We consider an $A$-body system containing the same number $n=A/n_f$ of 
$\sigma$ particles. We assume that the system is symmetric for the exchange of 
species, that is, single-particle orbitals are
occupied by all species of particles with an equal weight. Then  
the density matrix is diagonal with respect to $\sigma$, and 
all species of particles have the same occupation probability 
$\rho_{\sigma l}=\rho_l$ independent from $\sigma$. 
We consider the one-body density matrix and operator in the reduced space with the dimension $n$
and  define the entanglement measures, $S$, $K$, and $K_\gamma$, 
for a species of particles by using the reduced matrix of the one-body density.
The total entanglement entropy $S_{\textrm{total}}$
and the total numbers $K_{\textrm{total}}$ and $K_{\gamma,\textrm{total}}$
are given by the measures for each species 
as $S_\textrm{total}=n_f S$, $K_{\textrm{total}}=K$, and 
$K_{\gamma,\textrm{total}}=K_{\gamma}$. 
In this paper, we discuss $S$, $K$, and $K_\gamma$ for a species of particles.

For simplicity, $A$ particles are assumed to 
stay on sites in a space. The number of available sites (single-particle states) is $m$
and we use the label $k_j$ for the $j$th site. 
Let us first consider a system of $A=n_f$ particles. 
If an $A$-body state is an ideal state of independent particles,
the wave function can be written by a simple product
of single-particle wave functions 
\begin{eqnarray}
\Psi(1,2,\ldots,n_f)=\psi(1)\psi(2)\cdots\psi(n_f),\\
\psi(i)=\sum_{k=k_1,k_2,\ldots,k_{m}} c(k)  \phi_{k}(i).
\end{eqnarray}
For instance, $c(k)$ is constant as $c(k)=1/\sqrt{m}$ for a free gas state.
For such an non-entangled state, $S=0$, $K-1=0$, and $K_\gamma-1=0$
because the one-body density operator is given as 
$\hat \rho_{\Psi}=|\psi\rangle \langle\psi|$
and satisfies $\hat\rho_{\Psi}^2=\hat\rho_{\Psi}$.
Another example is  a ``localized cluster'' system of a cluster, where 
all particles are localized at one site $k_j$ to form a composite particle (a cluster) at 
$k_j$. The wave function is given as
\begin{equation}
\Psi(1,2,\ldots,n_f)=\prod_{i=1}^{n_f} \phi_{k_j} (i).
\end{equation}
This wave function for a localized clusters also has zero measures, 
 $S=0$, $K-1=0$, and $K_\gamma-1=0$, because 
the one-body density operator is given as 
$\hat \rho_{\Psi}=| \phi_{k_j}\rangle \langle  \phi_{k_j}|$
and satisfies again $\hat\rho_{\Psi}^2=\hat\rho_{\Psi}$.
Namely, the localized cluster wave function is a non-entangled state.

We consider a state of a delocalized cluster in a strong correlation limit, 
\begin{equation}
\Psi(1,2,\ldots,n_f)=\frac{1}{\sqrt{m}}\left\{
\prod_{i=1}^{n_f} \phi_{k_1} (i)+\prod_{i=1}^{n_f} \phi_{k_2} (i)+\cdots
\prod_{i=1}^{n_f} \phi_{k_{m}} (i)\right\},
\end{equation}
where the composite particle composed of $A=n_f$ particles 
moves freely in the entire space with an equal probability $\frac{1}{{m}}$. 
This is a highly entangled (strongly correlated) state, where, if a particle is observed 
at a certain site, all other particles are always observed at the same site. 
This is a strong coupling limit and an example of a delocalized cluster.
The one-body density operator,
\begin{equation}
\hat \rho_\Psi= \sum_{j=1}^{m} \frac{1}{m}|k_j\rangle \langle k_j|,
\end{equation}
corresponds to the Schmidt decomposition with the common
Schmidt coefficients, $1/m$.
We get $S=\log m$, $K=m$, and $K_\gamma=m$.
In general, if eigen values of $\rho_l$ are constant 
$\rho_l=1/m_V$ ($l=1,\ldots,m_V$) for a given number ($m_V$) 
of states and $\rho_l=0$ for $l>m_V$, we get
\begin{eqnarray}
S&=&\log{m_V}, \qquad \textrm{e}^S= m_V,\\
K&=&m_V,\\
K_\gamma&=&m_V.
\end{eqnarray} 
It indicates that $\textrm{e}^S$, $K$, and $K_\gamma$ equal to 
the number $m_V$ of states involved in the Schmidt decomposition. 
$m_V$ is regarded as the 
effective volume size. Note that, for a Slater determinant, 
$\textrm{e}^S$, $K$, and $K_\gamma$ equal to 1 indicating that the effective volume size is one
which can not be decomposed.

Next we consider an $A$-particle system of $n=A/n_f$ clusters. 
Here $n$ is the number of clusters (composite particles) 
formed by $n_f$ constituent Fermions.
For a state of $n$ localized clusters at $n$ sites $k={k_{j_1},\ldots,k_{j_n}}$, 
the wave function is given as,
\begin{eqnarray}
\Psi(1,2,\ldots,A)&=&(n!)^{-n_f/2}{\cal A}\left\{
\prod_{h=0}^{n-1}  \phi_{k_{j_1}}(hn_f+1) \cdots \phi_{k_{j_1}}(hn_f+n_f) 
\right\}\nonumber\\
&=&(n!)^{-n_f/2}{\cal A}\left\{ \phi_{k_{j_h}}(1) \cdots \phi_{k_{j_1}}(n_f)
\cdots\cdots
\phi_{k_{j_n}}(A-n_f+1) \cdots\phi_{k_{j_n}}(A)\right\}.
\end{eqnarray}
Since the occupation probability is $\rho_l=1$ for occupied single-particle states
$k_{j_1},\ldots,k_{j_n}$ and $\rho_l=0$ for unoccupied single-particle states, 
the system has the zero value of the measures, 
$S=0$, $K-1=0$, and $K_\gamma-1=0$.

Let us consider a state of $n$ clusters in the delocalized limit 
where all clusters are delocalized and move freely in a volume size 
$m_V$ like a gas. For this state of delocalized clusters, 
the occupation probability is $\rho_l=n/m_V$ for $l=1,\ldots,m_V$.
$m_V$ should not be less than $n$ because of the Pauli principle of $n$ Fermions.
The measures of this delocalized cluster system are,  
\begin{eqnarray}
S&=&n \log \frac{m_V}{n}, \qquad \textrm{e}^{S/n}= \frac{m_V}{n},\label{eq:delocalized1}\\
K&=&\frac{m_V}{n},\label{eq:delocalized2}\\
K_\gamma&=&\frac{m_V}{n}.\label{eq:delocalized3}
\end{eqnarray} 
It indicates that $K$ and $K_\gamma$ are consistent with $\textrm{e}^{S/n}$. 
$n\textrm{e}^{S/n}$, $nK$, and $nK_\gamma$ equal to the number $m_V$ of the states involved in the
Schmidt decomposition and they estimate the effective volume size of the delocalization of clusters. 
For the case of $m_V=n=A/n_f$, all $m_V$ single-particle states are completely occupied by $A$ particles
and clusters can not move at all. The state is equivalent to the localized cluster, and it has 
zero value, $S=0$, $K-1=0$, and $K_\gamma-1=0$, of entanglement measures. 
In case $m_V$ is larger than $n$, the measures, $S$, $K-1$, and $K_\gamma-1$,
become positive indicating that the delocalization of clusters occurs and the system
becomes an entangled state.
In the case that $m_V$ is much larger than $n$, the system corresponds to 
a low-density cluster gas and it is a highly entangled state.

Finally, we consider the case of a partial delocalization 
that $n-1$ clusters are localized to form a core and only the last cluster is delocalized. 
This situation corresponds to an $\alpha$ cluster moving 
almost freely around a core nucleus. 
In this partially localized case of a delocalized cluster around a core, 
the occupation probability is $\rho_l=n/m_V$ for  $l=1,\ldots,m_V$
and $\rho_l=1$ for the $n-1$ single-particle states occupied by constituent particles of the core.
Then,  the measures of this partially delocalized system are 
\begin{eqnarray}
S&=& \log {m_V}, \qquad \textrm{e}^{S/n}= m_V^{1/n},\label{eq:partial-delocalize1}\\
K&=&\frac{nm_V}{(n-1)m_V+1},\label{eq:partial-delocalize2}\\
K_\gamma&=&\frac{(n-1)m^\gamma_V+m_V}{(n-1)m^\gamma_V+1}.
\label{eq:partial-delocalize3}
\end{eqnarray} 
In the low-density limit of the large $m_V$, we get
\begin{eqnarray}
K&\rightarrow&\frac{n}{n-1},\label{eq:partial-delocalize-limit1}\\
K_\gamma&\rightarrow&m^{1-\gamma}_V.\label{eq:partial-delocalize-limit2}
\end{eqnarray} 

Let us compare the fully delocalized cluster state and 
the partially delocalized cluster state in the large $m_V$ limit 
(the large volume size limit).
The former state corresponds to a dilute cluster gas, where 
all clusters are delocalized  moving freely like a gas, and is a
highly entangled state. For the fully delocalized cluster state, 
all the entanglement measures are sensitive to $m_V$ as given in 
Eqs.~\eqref{eq:delocalized1}, \eqref{eq:delocalized2}, and \eqref{eq:delocalized3}. That is, $\textrm{e}^{S/n}$,  $K$, and $K_\gamma$
equal to $m_V/n$ and equivalently good indicators to measure the
delocalization of clusters.
However, for the latter case of the partial delocalization, 
which corresponds to a delocalized cluster around a core, 
$\textrm{e}^{S/n}$, $K$, and $K_\gamma$ are not equivalent but 
show different dependences on $m_V$. 
As clearly shown in  Eqs.~\eqref{eq:partial-delocalize1}, \eqref{eq:partial-delocalize-limit1},
and \eqref{eq:partial-delocalize-limit2},
$\textrm{e}^{S/n}$ and $K_\gamma$ increases as the $m_V$ increases, but $K$ becomes 
constant and does not depend on $m_V$ in the large $m_V$ limit.
It indicates that $S$ and $K_\gamma$ can be useful measures sensitive to
the partial delocalization in subsystem, 
but $K$ is insensitive to the partial delocalization. The reason for the insensitivity of $K$
is the significant contribution from fully occupied single-particle states 
with $\rho_l=1$ in the denominator of the definition of the $K$ number, 
which makes the contribution from the delocalized part minor. 
From another point of view, it is found that
$K$ can be a good probe to clarify whether the delocalization of clusters 
occurs in the entire system or not. 
$m_V$ dependences of $\textrm{e}^{S/n}$ and $K_\gamma$ are 
the powers of $1/n$ and $1-\gamma$, respectively. 
More generally, for the partially delocalized system that is
composed of $n_g$ delocalized clusters and $n_0$ localized clusters,
the $m_V$ dependences of $\textrm{e}^{S/n}$ and $K_\gamma$ in the large $m_V$ limit 
are the powers of $n_g/n$ and $1-\gamma$, respectively. 
Here, $n_g$ and $n_0$ are the numbers of delocalized and localized clusters,
respectively, and $n_0+n_g=n$.
This means
that, in the case $n_g/n > 1-\gamma$ of a small fraction of delocalized clusters,  
$K_\gamma$ is more sensitive to the delocalization of clusters than $\textrm{e}^{S/n}$,
whereas, 
in the case of $n_g/n \approx 1-\gamma$,  the $K_\gamma$ number has 
the $m_V$ dependence similar to $\textrm{e}^{S/n}$. 

\section{Application to 1D nuclear systems of $\alpha$ clusters}\label{sec:1D}
In the present paper, we use the delocalized cluster wave functions in 1D
for the linear-chain $n\alpha$ states
which are investigated in the previous paper \cite{enyo-ee}. 
We also adopt the $\alpha+(2\alpha)$  wave functions 
for a state of an $\alpha$ cluster around a $2\alpha$ core.
We analyze the entanglement measures ,
$S$, $K$, and $K_{\log 2}$, and also their spatial distributions, and 
discuss the system size dependence of these measures.

\subsection{Localized and delocalized $\alpha$ cluster wave functions in 1D}
We here briefly explain the adopted model wave functions for
(de)localized cluster states in 1D. More details of the model wave functions 
are described in the previous paper \cite{enyo-ee}.
We consider intrinsic wave functions of the linear-chain  $n\alpha$-cluster states
aligned to the $x$ axis. It means that the (de)localization of $\alpha$ clusters 
are defined for $\alpha$-cluster motion along the $x$ axis. 

For a localized $n\alpha$-cluster wave function,
we use the BB wave function \cite{brink66} as,
\begin{eqnarray}
\Phi^{n\alpha}_{\rm BB}(\bvec{R}_1,\ldots,\bvec{R}_{n})&=&\frac{1}{\sqrt{A!}}
{\cal A}\left[ \psi^\alpha_{\bvec{R}_1}\cdots  \psi^\alpha_{\bvec{R}_{n}}\right],\\
\psi^\alpha_{\bvec{R}_i}&=&\phi^{0s}_{\bvec{R}_i}\chi_{p\uparrow}
\phi^{0s}_{\bvec{R}_i}\chi_{p\downarrow}\phi^{0s}_{\bvec{R}_i}\chi_{n\uparrow}
\phi^{0s}_{\bvec{R}_i}\chi_{n\downarrow},\\
\phi^{0s}&=&(\pi b^2)^{-3/4}\exp\left[ -\frac{1}{2b^2}(\bvec{r}-\bvec{R}_i)^2\right].
\end{eqnarray}
$\psi^\alpha_{\bvec{R}_i}$ is the four-nucleon wave function of the 
$i$th $\alpha$ cluster expressed by the $(0s)^4$ harmonic oscillator (ho) shell-model configuration 
localized around the spatial position $\bvec{R}_i$. $\chi$ is the spin-isospin part and 
$\phi^{0s}$ is the spatial part of the single-particle wave function.
For 1D cluster states, the position parameter $\bvec{R}_i$
is set to be $\bvec{R}_i=(R_i,0,0)$, and the 1D BB wave function is expressed as
$\Phi^{n\alpha}_{\rm BB}(R_1,\ldots,R_{n})$.
The parameter $b$ for the $\alpha$-cluster size
is chosen to be $b=1.376$ fm same as in Ref.~\cite{Suhara:2013csa}.
Note that a single BB wave function is a localized cluster wave function
written by a Slater determinant of 
single-particle wave functions and it has exactly zero values of the measures, $S=0$, $K-1=0$, and $K_\gamma-1=0$.
General wave functions for 1D $n\alpha$ systems 
can be written by linear combination of  BB wave functions
$\Phi^{n\alpha}_{\rm BB}(R_1,\ldots,R_{n})$.

For a delocalized cluster wave function, we use the 1D-THSR wave functions
of $n\alpha$, which have been 
proposed by Suhara {\it et al.} to describe the 
linear-chain $3\alpha$- and $4\alpha$-cluster states 
in $^{12}$C and $^{16}$O systems \cite{Suhara:2013csa}.
The 1D-THSR wave functions are given by linear combination of  BB wave functions 
with a Gaussian weight as,
\begin{eqnarray}
\Phi^{n\alpha}_\textrm{1D-THSR}(\beta)=\int dR_1 \cdots dR_{n}\exp\left\{ 
-\sum^{n}_{i=1} \frac{R^2_i}{\beta^2}
\right\}\Phi^{n\alpha}_{\rm BB}(R_1,\ldots,R_{n}).
\end{eqnarray}
If the antisymmetrization is ignored, the $\Phi^{n\alpha}_\textrm{1D-THSR}(\beta)$ expresses the 
$n\alpha$ state where all $\alpha$ clusters are confined in the 
$y$ and $z$ directions while they move in the $x$ direction in the 
Gaussian orbit with the range parameter $\beta$. $\beta$
corresponds to the system size of the 1D $n\alpha$ state.
In case of the system size $\beta$ is as small as or smaller than the $\alpha$-cluster size $b$,   
the 1D-THSR wave function is approximately equivalent to a localized 
cluster wave function given by a Slater determinant
because of the antisymmetrization effect. 
As $\beta$ increases, the delocalization of $\alpha$ clusters occurs.
When $\beta$ is large enough compared with the $\alpha$-cluster size $b$, 
the system goes to a dilute 1D $\alpha$-cluster gas 
where $n$ $\alpha$ clusters move almost freely like a gas in the $x$ direction.

In the practical calculation, the $R_i$ integration is approximated by 
summation on mesh points in a finite box as done in the previous paper. 
For $2\alpha$, 3$\alpha$, and 4$\alpha$ systems, 
we make a correction of 
the total c.m.m. to eliminate a possible artifact from $\beta$ dependence 
in the total c.m.m. as described in the previous paper.
It means that the 1D-THSR wave function of $1\alpha$ without the c.m.m. correction 
expresses a system of an $\alpha$ cluster bound in an external field,
which is not a realistic state of an isolate nucleus, 
whereas those of $2\alpha$,  $3\alpha$, and $4\alpha$ with the c.m.m. correction correspond to
self bound $n\alpha$ systems with linear-chain structures predicted 
in nuclear states such as excited states of $^{12}$C and $^{16}$O.
In this paper, we discuss the entanglement  measures for $1\alpha$ with no c.m.m. correction (nc)
and those for $2\alpha$,  $3\alpha$, and $4\alpha$ with the c.m.m. correlation. We also show the enganlement
entropy  for  $2\alpha$ with no c.m.m. correction, just for comparison. 

We also use the 1D-THSR wave function of $\alpha+(2\alpha)$ for 
the case of an $\alpha$ cluster around the $2\alpha$ core,
\begin{eqnarray}\label{eq:a-2a}
\Phi^{\alpha\textrm{-}(2\alpha)}_\textrm{1D-THSR}(\beta)&=&\int dR_1 
\exp\left\{ -\frac{R_1^2}{\beta^2}
\right\}\Phi^{3\alpha}_{\rm BB}(R_1,R_2=+\varepsilon,R_3=-\varepsilon), 
\end{eqnarray}
where the $2\alpha$ core is located at the origin and 
an $\alpha$ cluster is distributed around the core 
with a Gaussian weight.
When $\beta$ is large enough compared with the $\alpha$-cluster size $b$, 
the wave function describes a delocalized $\alpha$ cluster around 
the $2\alpha$ core at the origin. 
This wave function is associated with 
the partially delocalized cluster wave function in the toy model discussed previously.
We use a small value 
of $\varepsilon=0.02$ fm to describe
the $2\alpha$ core almost equivalent to the h.o. $(0s)^4(0p_x)^4$ configuration.
This wave function has been originally introduced in previous paper 
to describe the $\alpha+^{16}$O cluster states in $^{20}$Ne. 

We calculate numerically the one-body density matrices 
for these 1D cluster wave functions, and 
obtain the measures, $S$, $K$, and $K_{\log 2}$ from the eigen values $\rho_l$ 
of the one-body density matrices. The detailed method of the practical calculation is described 
in the previous paper.
Spatial distributions, $s(\bvec{r})$, $\kappa(\bvec{r})$, and 
$\kappa_{\gamma}(\bvec{r})$ and the density distribution $\rho(\bvec{r})$  
are integrated out along $y$ and $z$ axes, and  
we get distributions, 
$s(x)$, $\kappa(x)$, $\kappa_\gamma(x)$, and $\rho(x)$, projected onto the $x$ axis.

\subsection{$n\alpha$ cluster states}

\begin{figure}[htb]
\begin{center}
	\includegraphics[width=7.5cm]{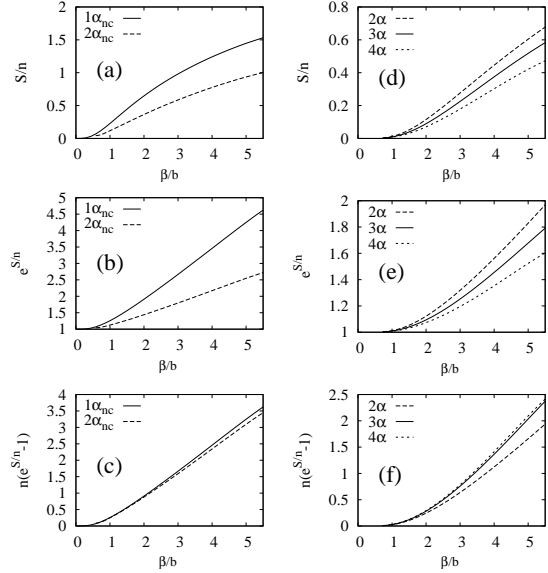} 	
\end{center}
  \caption{System size $\beta/b$ dependence of 
the entanglement entropy $S$ and $\textrm{e}^{S/n}$ in the 1D-THSR wave functions
of $n\alpha$.  $n(\textrm{e}^{S/n}-1)$ is also shown. 
The 1D-THSR wave functions with the total c.m.m. correction are used for 
$2\alpha$, $3\alpha$, and $4\alpha$ systems, 
and that with no c.m.m. correction is used for 
the $1\alpha$ system ($1\alpha_{\rm nc}$). The results of the 1D-THSR wave function
with no  c.m.m. correction for $2\alpha$ system ($2\alpha_{\rm nc}$) are also shown for
comparison. 
\label{fig:s-1a}}
\end{figure}

We analyze $S$, $K$, and $K_{\log 2}$ of 
the 1D $n\alpha$ states written by the 1D-THSR wave functions, $\Phi^{n\alpha}_\textrm{1D-THSR}(\beta)$.
Figure \ref{fig:s-1a} shows the system size dependence of the entanglement entropy of 
the $1\alpha$ state with no c.m.m. correction ($1\alpha_{\rm nc}$), 
and $2\alpha$, $3\alpha$, and $4\alpha$ states with the c.m.m. correction, as well as that of a 
$2\alpha$ state with no c.m.m. correction ($2\alpha_{\rm nc}$), 
$S/n$, $\textrm{e}^{S/n}$, and $n (\textrm{e}^{S/n}-1)$ are plotted as functions of the dimensionless system size 
$\beta/b$. 
In the $\beta/b=0$ limit, the entanglement entropy equals zero, because the 1D-THSR wave functions are
equivalent to localized $n\alpha$ cluster wave functions in this limit.  As the $\beta/b$
increases, the delocalization of clusters occurs and the entanglement entropy is generated.
$\textrm{e}^{S/n}$ for the $1\alpha_{\rm nc}$ state increases 
almost linearly to $\beta/b$ as expected from a naive picture that a cluster moves freely in a finite volume, which is decomposed into $m_V\sim\beta/b$ states by the 
quantum decoherence in the one-body density matrix.
Similarly to the $1\alpha_{\rm nc}$ state, $S$ 
for $2\alpha$, $3\alpha$, and $4\alpha$ states increases 
with the increase of the system size, indicating 
that the entanglement entropy is generated as the delocalization of clusters is enhanced. 
$\textrm{e}^{S/n}$ for $2\alpha$, $3\alpha$, and $4\alpha$ states also shows 
almost linear dependence to $\beta/b$. 
The $n(\textrm{e}^{S/n}-1)$ plots in Figs.~\ref{fig:s-1a}(c) and \ref{fig:s-1a}(d) show 
a good correspondence of the $\beta$ dependence of the 
entanglement entropy between $1\alpha_{\rm nc}$ and $2\alpha_{\rm nc}$, and that 
between $2\alpha$, $3\alpha$, and $4\alpha$ states. The values of 
$n(\textrm{e}^{S/n}-1)$ for $2\alpha$, $3\alpha$, and $4\alpha$ are
relatively smaller compared with 
those for $1\alpha_{\rm nc}$ and $2\alpha_{\rm nc}$ because of the 
c.m.m correction performed for the $2\alpha$, $3\alpha$, and $4\alpha$ states.

\begin{figure}[htb]
\begin{center}
	\includegraphics[width=11.25cm]{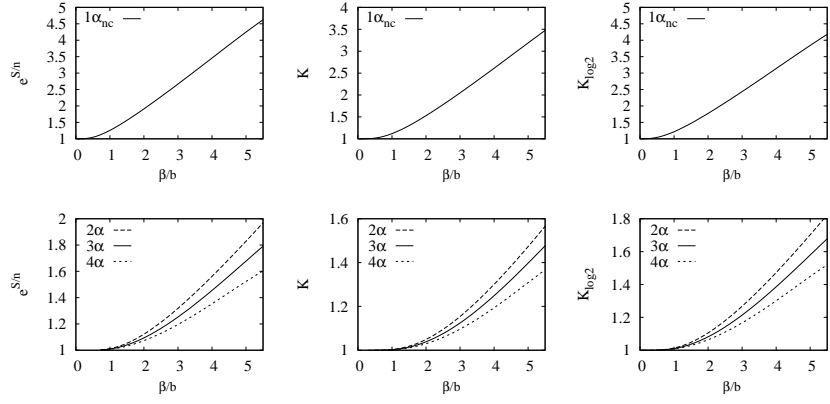} 	
\end{center}
  \caption{Comparison of system size $\beta/b$ 
dependences between $\textrm{e}^{S/n}$, $K$, and $K_{\log 2}$ in the 1D-THSR wave functions
of $n\alpha$. \label{fig:s.1a2a3a4a}}
\end{figure}

Let us compare other measures, $K$ and $K_{\log 2}$, with the entanglement entropy.
Figure \ref{fig:s.1a2a3a4a} shows the system size dependence of $\textrm{e}^{S/n}$, $K$, and $K_{\log 2}$
of $1\alpha_{\rm nc}$, $2\alpha$, $3\alpha$, and $4\alpha$ states. $K$ and $K_{\log 2}$ shows 
the $\beta/b$ dependence quite similar to that of $\textrm{e}^{S/n}$ except for global normalization factors.
This result indicates that the entanglement entropy, $K$, and $K_{\log 2}$, can be approximately equivalent measures 
for the cluster delocalization of $n\alpha$ states. It is consistent with the naive expectation from  
the analysis of the delocalized cluster states in the toy model discussed in the previous section.
It means that $n\textrm{e}^{S/n}$, $nK$, and $nK_{\log 2}$ estimate the number of the states involved 
in the Schmidt decomposition as shown in Eqs.~\eqref{eq:delocalized1},
\eqref{eq:delocalized2}, and \eqref{eq:delocalized3}.

\begin{figure}[htb]
\begin{center}
	\includegraphics[width=15cm]{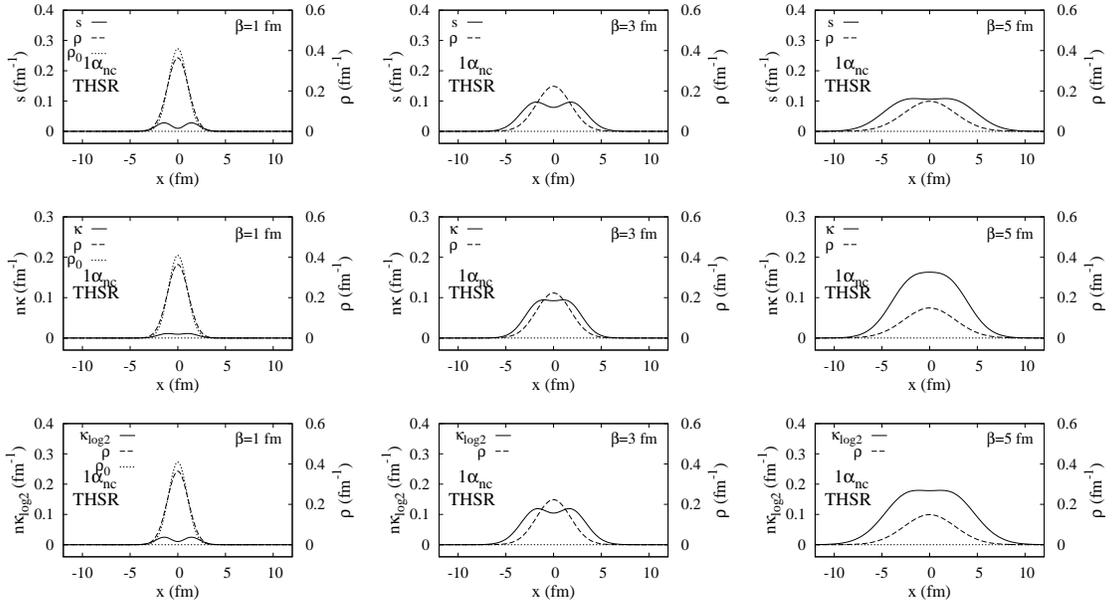} 	
\end{center}
  \caption{Spatial distributions, $s(x)$, $\kappa(x)$, and $\kappa_{\log2}(x)$, of $S$, 
$K-1$, and $K_{\log2}-1$ in the 1D-THSR wave functions of $1\alpha_{\rm nc}$ with 
$\beta=1$ fm, 3 fm, and 5 fm. The corresponding dimensionless system sizes are
$\beta/b=0.73$, 2.18, and 3.63. The density distribution $\rho(x)$ is also shown.
\label{fig:sd-1a}}
\end{figure}

\begin{figure}[htb]
\begin{center}
	\includegraphics[width=15cm]{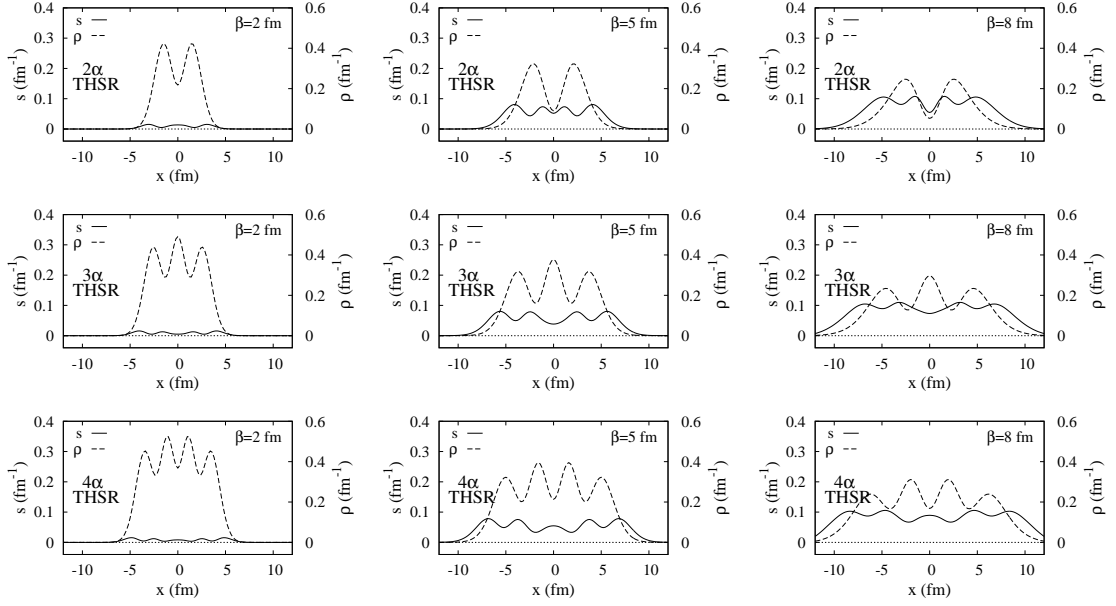} 	
\end{center}
  \caption{Spatial distribution, $s(x)$, of the entanglement entropy $S$ 
in the 1D-THSR wave functions of $2\alpha$, $3\alpha$, and $4\alpha$ with 
$\beta=2$ fm, 5 fm, and 8 fm. The corresponding dimensionless system sizes are
$\beta/b=1.45$, 3.63, and 5.81. The density distribution $\rho(x)$ is also shown.
\label{fig:sd-2a3a4a}}
\end{figure}

We also compare the spatial distributions of these measures, $s(x)$, $\kappa(x)$, and $\kappa_{\log 2}(x)$ of $1\alpha_{\rm nc}$ in Fig.~\ref{fig:sd-1a}. 
The density distribution is also shown. 
 As described previously, $s(x)$, $\kappa(x)$, and $\kappa_{\log 2}(x)$ reflect 
spatial distributions of the important single-particle states that give non-zero contributions 
to total measures $S$, $K-1$, $K_{\log 2}-1$, respectively. Note that the shape, in particular, 
the spatial broadness of distributions is of importance.
but their global scaling (normalization) is not so meaningful. 
It is found that 
$s(x)$ and $\kappa_{\log 2}(x)$ show quite similar distributions to each other. 
They are more broadly distributed than the density distribution 
indicating that $S$ and $K_{\log 2}-1$ are generated in low-density regions but relatively 
suppressed in high-density regions. Differently from $s(x)$ and $\kappa_{\log 2}(x)$, 
the enhancement in low-density regions and the suppression in high-density regions
are relatively weak in $\kappa(x)$. As a result, the spatial extent of $\kappa(x)$ is not as remarkable as
that of  $s(x)$ and $\kappa_{\log 2}(x)$. This difference in the spatial distributions of 
$S$, $K-1$, $K_{\log 2}-1$ comes from the different weight functions $W(\rho)$, namely, 
$S$ and $K_{\log 2}-1$ are 
relatively sensitive to single-particle states with low occupation probability compared with $K-1$
as already shown in Fig.~\ref{fig:dens}.
The distributions $s(x)$, $\kappa(x)$, and $\kappa_{\log 2}(x)$ for the $2\alpha$, $3\alpha$, and $4\alpha$ states 
are shown in Fig.~\ref{fig:sd-2a3a4a} as well as density distribution. 
Similarly to the $1\alpha_{\rm nc}$ state, 
$s(x)$ and $\kappa_{\log 2}(x)$ 
are more broadly distributed than the density distribution and also slightly broader than $\kappa(x)$,
indicating again that $S$ and $K_{\log 2}-1$ are generated 
in low-density regions but suppressed in high-density regions.

In the present result, we find that the $S$, $K$ and $K_{\log 2}$ can be 
useful measures to estimate the entanglement caused by the delocalization of clusters 
in the 1D $n\alpha$ states
given by the 1D-THSR wave functions.
As the system size increases, the delocalization of clusters develops and
non-zero $S$, $K-1$, and $K_{\log 2}-1$ are generated. 
In the spatial distributions of these entanglement measures, significant contributions 
come from low-density regions than high-density regions.
Quantitatively, the Schmidt number ($K$) is less sensitive to low-occupation 
probability single-particle states than the entanglement entropy ($S$) and the $K_{\log 2}$ number 
resulting in the less broad distribution of $\kappa(x)$ than 
$s(x)$ and $\kappa_{\log 2}(x)$, which are remarkably broader
than the density distribution.
 
\begin{figure}[htb]
\begin{center}
	\includegraphics[width=15cm]{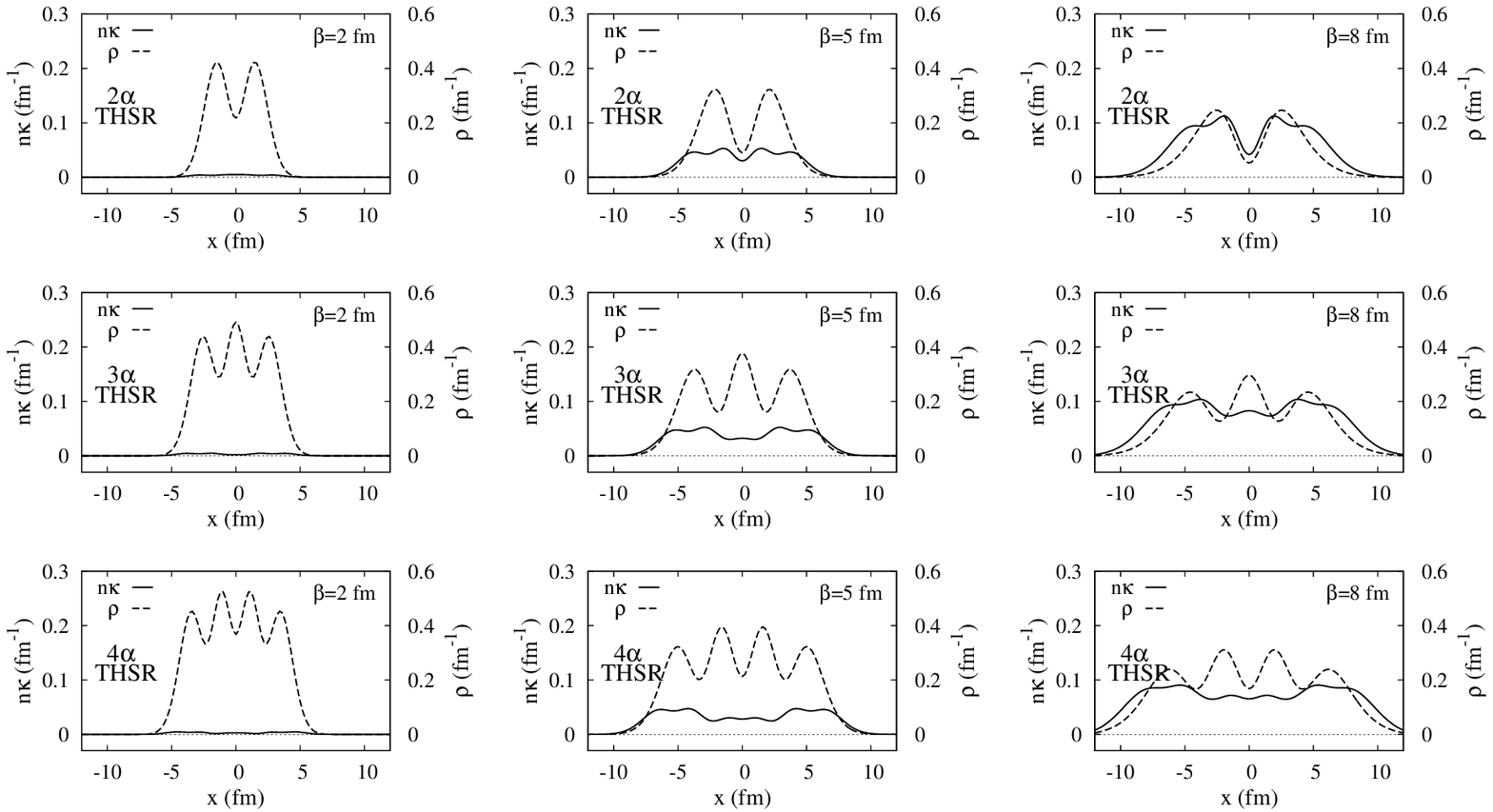} 	
\end{center}
  \caption{Same as Fig.~\ref{fig:sd-2a3a4a} but 
spatial distribution, $\kappa(x)$, for $K-1$.  
\label{fig:k1d-2a3a4a}}
\end{figure}

\begin{figure}[htb]
\begin{center}
	\includegraphics[width=15cm]{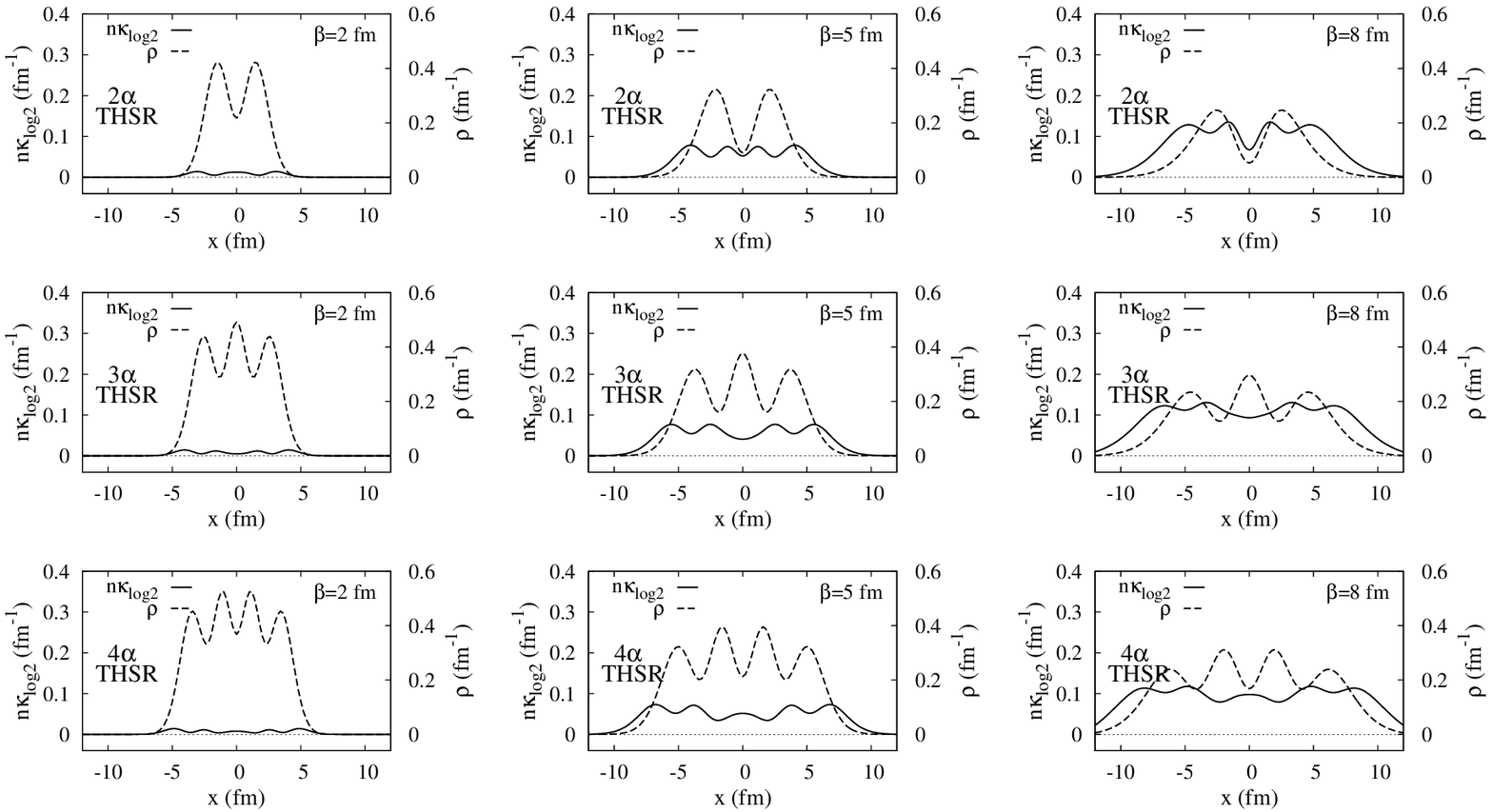} 	
\end{center}
  \caption{Same as Fig.~\ref{fig:sd-2a3a4a} but 
spatial distribution, $\kappa_{\log 2}(x)$, for  $K_{\log 2}-1$.  
\label{fig:k2d-2a3a4a}}
\end{figure}

\subsection{an $\alpha$ cluster around a $2\alpha$ core}
We investigate $S$, $K$, and $K_{\log 2}$ 
of $\alpha+(2\alpha)$ cluster wave functions. We use the 1D-THSR wave function 
$\Phi^{\alpha\textrm{-}(2\alpha)}_\textrm{1D-THSR}(\beta)$ in Eq.~\eqref{eq:a-2a} 
for the $\alpha+(2\alpha)$ states associated with the partially 
delocalized cluster wave function. 
We also adopt the parity projected 
BB wave function with the fixed 
$\alpha$-$(2\alpha)$ distance $d$ as used in the previous paper
as, 
\begin{eqnarray}
\Phi^{\alpha\textrm{-}(2\alpha),+}_{\rm BB}(d)&=&
(1+\hat{P}_r)\Phi^{3\alpha}_{\rm BB} 
(R_1=d,R_2=+\varepsilon,R_3=-\varepsilon)\nonumber\\
&=&\Phi^{3\alpha}_{\rm BB} 
(R_1=d,R_2=+\varepsilon,R_3=-\varepsilon)\nonumber\\
&&+\Phi^{3\alpha}_{\rm BB}(R_1=-d,R_2=+\varepsilon,R_3=-\varepsilon),
\end{eqnarray}
where $\hat{P}_r$ is the parity transformation operator.
Note that $\Phi^{\alpha\textrm{-}(2\alpha),+}_{\rm BB}(d)$ 
is given by the linear combination of two Slater determinants.
This corresponds to the symmetry restored state where 
the parity symmetry is broken in the intrinsic state because of the cluster development.

\begin{figure}[htb]
\begin{center}
	\includegraphics[width=15cm]{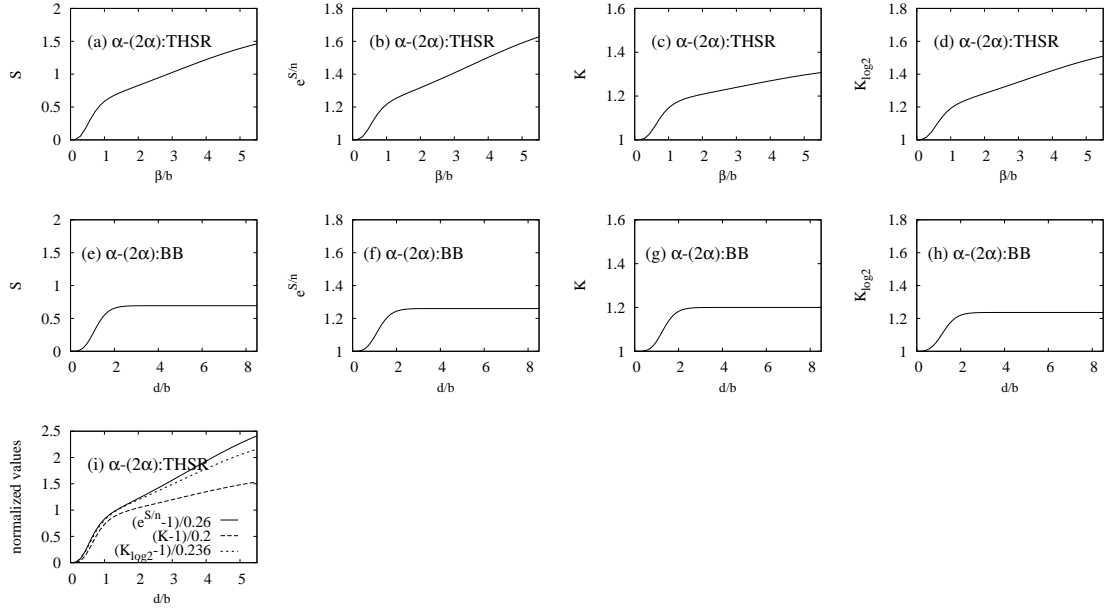} 	
\end{center}
  \caption{(a)(b)(c)(d) System size $\beta/b$ dependence of 
$S$, $\textrm{e}^{S/n}$, $K$, and $K_{\log2}$ 
in the 1D-THSR wave function of $\alpha+(2\alpha)$,  and  
(e)(f)(g)(h) $d/b$ dependence of $S$,  $\textrm{e}^{S/n}$, $K$, and $K_{\log2}$ 
in the parity-projected BB wave function of  $\alpha+(2\alpha)$. 
(i) Scaled measures, 
$(\textrm{e}^{S/n}-1)/0.26$, $(K-1)/0.2$, $K_{\log 2}/0.236$, in the 1D-THSR wave function of $\alpha+(2\alpha)$, normalized to the large $d$ limit values of the parity-projected BB wave function.
\label{fig:s.2a-a}}
\end{figure}

Figure \ref{fig:s.2a-a} shows $S$, $K$, and $K_{\log 2}$ as well as
$\textrm{e}^{S/n}$ for the 1D THSR wave function 
and the parity-projected BB wave function of the $\alpha+(2\alpha)$ system.
Looking at the result of the parity-projected BB wave function, we find that, as $d$ increases 
and the parity symmetry is broken in the intrinsic wave function, 
non-zero values of  $S$,  $K-1$, and $K_{\log 2}$ are generated in the projected state 
even though the intrinsic wave function before the parity projection is 
the localized cluster wave function.
When $d$ is enough large, $\hat{P}_r\Phi^{3\alpha}_{\rm BB}$ 
becomes independent to $\Phi^{3\alpha}_{\rm BB}$, and we get 
$S\rightarrow \log 2=0.693$, $K-1\rightarrow 1/5=0.2$, $K_{\log 2}-1\rightarrow 0.236$
from the eigen values of the one-body density matrix, 
$\rho_1=\rho_2=1$ and $\rho_3=\rho_4=1/2$. 
It means that 
non-zero values of $S$,  $K-1$, and $K_{\log 2}-1$ are generated 
by the symmetry breaking and restoration. 
Let us turn to the result of the 1D-THSR wave function. 
In the $\beta/b=0$ limit,  $S$,  $K-1$, and $K_{\log 2}-1$ are zero. 
 In the $\beta/b \lesssim 1$ region,  
$S$,  $K$, and $K_{\log 2}$ increases 
rapidly with the increase of $\beta/b$ 
because of the symmetry breaking and restoration as seen in the parity-projected
BB wave function. 
In the $\beta/b \gtrsim 1$, $S$,  $K$, and $K_{\log 2}$ increases 
gradually as the system size increases indicating that 
the delocalization of cluster develops in this region. 

As discussed in the previous section, the Schmidt number, $K$, 
should be less sensitive to the delocalization in the partially delocalized cluster states.
To see the sensitivity of $\textrm{e}^{S/n}$, $K$, and $K_{\log 2}$  to the delocalization, 
we show in Fig.~\ref{fig:s.2a-a}(i) the scaled measures, 
$(\textrm{e}^{S/n}-1)/0.26$, $(K-1)/0.2$, $K_{\log 2}/0.236$,  in the 1D-THSR wave function of 
$\alpha+(2\alpha)$, 
which are normalized to the values of the large $d$ limit of 
the parity-projected BB wave function.
As expected from the analysis of the simple toy model, the result in Fig.~\ref{fig:s.2a-a}(i)
shows that
$K$ is not so sensitive to the delocalization of cluster in the $\alpha+(2\alpha)$ state,
whereas $\textrm{e}^{S/n}$ and $K_{\log 2}$ more strongly depend on the 
the system size in the $\beta/b \gtrsim 1$ region than $K$. 
The $\beta/b$ dependence of $K_{\log 2}$ is quite similar to that of $\textrm{e}^{S/n}$, 
maybe, because of the accidental coincidence of the $m^{1/n}_V$ dependences,
$\textrm{e}^{S/n}\propto m^{1/n}_V$ 
and $K_\gamma\propto m^{(1-\gamma)}_V$,  
for the partially delocalized cluster state discussed in the simple toy model
as $1/n=1/3$ and $1-\gamma=1-\log2=0.31$.

\begin{figure}[htb]
\begin{center}
	\includegraphics[width=10cm]{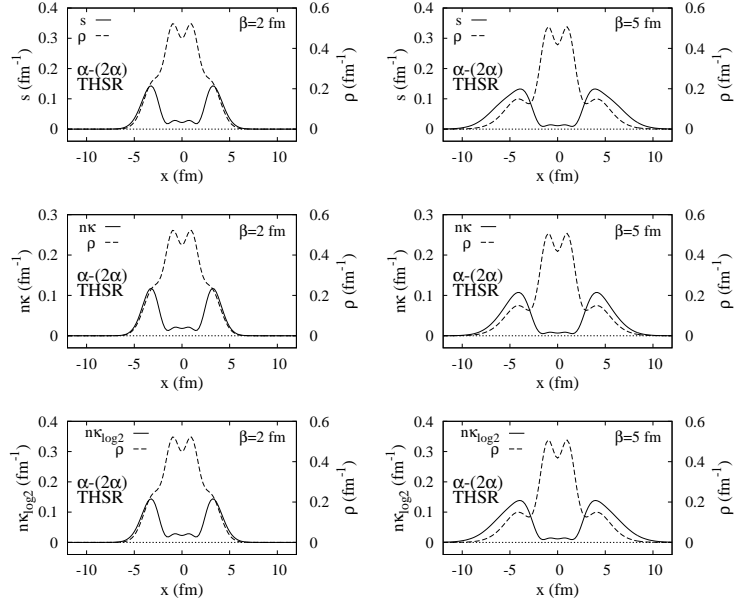} 	
\end{center}
  \caption{Spatial distributions, $s(x)$, $\kappa(x)$, and $\kappa_{\log2}(x)$, of $S$, 
$K$, and $K_{\log2}$ in the 1D-THSR wave functions of $\alpha+(2\alpha)$ with 
$\beta=2$ fm and 5 fm. The corresponding dimensionless system sizes are
$\beta/b=1.45$ and 3.63. The density distribution $\rho(x)$ is also shown.
\label{fig:sd-2a-a}}
\end{figure}

Figure \ref{fig:sd-2a-a} shows the distributions $s(x)$, $\kappa(x)$, 
and $\kappa_{\log 2}(x)$ of 
$S$, $K-1$, and $K_\gamma-1$ as well as the density distribution 
in the 1D-THSR wave functions 
of $\alpha+(2\alpha)$ for $\beta=2$ fm and 5 fm.
It is clear that distributions $s(x)$, $\kappa(x)$, 
and $\kappa_{\log 2}(x)$ are strongly suppressed in the $|x|\lesssim 2$ fm
because of the Pauli blocking effect from the $2\alpha$ core.
In the $\beta=5$ fm case, the delocalization occurs, and $S$ and $K_\gamma-1$ 
is generated in the low-density regions in particular at the long tail part.
Similarly to the distributions in $n\alpha$ states, the distribution 
$\kappa(x)$ for $K-1$ is not so enhanced in low-density regions as $s(x)$ and $\kappa_{\log 2}(x)$ 
because the Schmidt number $K$ is 
relatively less sensitive to single-particle states with 
low occupation probability compared with the entanglement entropy and the $K_{\log 2}$ number.

\section{Summary}\label{sec:summary}
We calculated the entanglement entropy ($S$) and the Schmidt number ($K$) 
defined by the one-body density matrix
in the 1D $\alpha$-cluster states to measure the entanglement caused 
by the delocalization of clusters in nuclear systems.
We also propose a new entanglement measure $K_\gamma$ with a generalized form of the 
Schmidt number. 

For the delocalized cluster states of $n\alpha$ given by the 
1D-THSR wave functions, $\textrm{e}^{S/n}$, $K$, and $K_{\log2}$ show
good correspondence indicating that 
the entanglement entropy, the Schmidt number, and the $K_{\log 2}$ number
are almost equivalent entanglement  measures to estimate 
the delocalization of a 1D gas of delocalized $n$ $\alpha$ clusters.
On the other hand, 
for the partially delocalized cluster state which contains a delocalized cluster
and a core composed of localized clusters, the Schmidt number is not sensitive to 
the delocalization of the cluster around the core, whereas the entanglement entropy and the $K_{\log 2}$ number
can be good indicators to measure the partial delocalization. 
In other words, the Schmidt number can be a good probe to clarify whether  
the delocalization occurs for all clusters in the entire system or not,
owing to the insensitivity to the partial delocalization.
We should comment that the Schmidt number corresponds to the 
the R\'enyi-2 entanglement entropy; $K=\textrm{e}^{S^{\rm Renyi}_2}$. 
It means that the equivalence in the fully delocalized cluster states and difference in the partially 
delocalized cluster states between $\textrm{e}^{S/n}$ and $K$ are nothing but 
those between the von Neumann entanglement entropy and R\'enyi-2 entanglement entropy.

In the present analysis of the 1D $\alpha$-cluster states, 
the $K_{\log 2}$ number shows similar features to the entanglement entropy.
It indicates that the $K_{\log 2}$ number can be an alternative measure to
the entanglement entropy to estimate the delocalization in both cases of the partially and fully 
delocalized cluster states. When the delocalized part is minor
compared with the core part, the $K_{\log 2}$ number is 
more sensitive to the partial delocalization than the entanglement entropy, and hence  
it is a promising measure.

We should point out that the calculation of 
the Schmidt number may be practically easier than those of 
the entanglement entropy and $K_\gamma$ with a non-integer $\gamma$ because 
the Schmidt number is given by $\textrm{Tr}{\rho}^2$ for which
the diagonalization of the one-body density matrix is not required.
This could be an advantage of the Schmidt number in large dimensional calculations  
because numerical errors might become a more serious problem in practical 
calculations of the entanglement entropy and the $K_\gamma$ number.

Recently, comparison between von Neumann and R\'enyi-2 
entanglement entropies have been discussed in various fields such as field theories \cite{Caputa:2014vaa}. 
The equivalence and difference between von Neumann and R\'enyi-2 entanglement entropies shown in the present study are general features and may be found in various quantum systems. 
The present study of entanglement measures in cluster wave functions of nuclear systems
may shed light on study of entanglement (correlation) in 
general quantum systems. 

\acknowledgments
The author would like to thank H.~Iida for helpful discussions.
The calculations of this work have been done using computers at Yukawa Institute for Theoretical
Physics, Kyoto University.
This work was supported by 
JSPS KAKENHI Grant No. 26400270.

\end{document}